# All linear optical devices are mode converters

**David A. B. Miller**

*Ginzton Laboratory, Stanford University, 348 Via Pueblo Mall, Stanford CA 94305-4088, USA*
*dabm@ee.stanford.edu*

**Abstract:** We show that every linear optical component can be completely described as a device that converts one set of orthogonal input modes, one by one, to a matching set of orthogonal output modes. This result holds for any linear optical structure with any specific variation in space and/or time of its structure. There are therefore preferred orthogonal "mode converter" basis sets of input and output functions for describing any linear optical device, in terms of which the device can be described by a simple diagonal operator. This result should help us understand what linear optical devices we can and cannot make. As illustrations, we use this approach to derive a general expression for the alignment tolerance of an efficient mode coupler and to prove that loss-less combining of orthogonal modes is impossible.

## 1. Introduction

In optics there is growing interest in understanding how we can make devices that convert specific kinds of inputs to specific kinds of outputs. Recent examples include mode converters [1-7], especially those that convert more than one different input to more than one different output [1-3], optical isolators based on time-varying dielectrics [8], and devices that unscramble or decode the outputs of multimode optical fibers [3, 9]. The growing capabilities of nanophotonic fabrication technologies mean a broad range of new or improved devices may be possible. Both for design and for understanding limits to what devices can be made, we therefore want a clear and simple approach to the mathematics of such devices.

In general with some optical device or scatterer, shining distinct, orthogonal beams on the input of the device does not lead to orthogonal output beams; there is no guarantee for linear operators that orthogonal inputs give orthogonal outputs. Here, however, we show that every linear optical device is completely describable as one that converts, one by one, from a specific set of orthogonal input modes to a specific set of orthogonal output modes. Every linear optical device is therefore completely describable as a mode converter; equivalently, it has a special set of orthogonal inputs that leads to orthogonal outputs; we can call these orthogonal sets the "mode converter" basis sets. These modes can also describe everything that this linear device can do.

Using the mathematics from this result, we expect to be able to understand some limits and possibilities in optical devices. In this paper, after proving the core result, we illustrate the use of this approach: First, we derive an expression for the misalignment tolerance of an efficient mode coupler, showing a simple and universal result and, second, we prove that loss-less beam combination of two orthogonal modes into one is impossible.

## 2. The mode converter basis set

*2.1 Device operator*

A linear optical component is a device on which we shine light beams and/or pulses and from which we get corresponding output light beams and/or pulses in a linear fashion. Such a linear device can always be described by some linear operator $\mathsf{D}$ that we can call the device operator. The mathematical function of $\mathsf{D}$ is therefore to take an input function $|\phi_I\rangle$ and to generate a corresponding output function $|\phi_O\rangle$, that is

$$|\phi_O\rangle = \mathsf{D}|\phi_I\rangle. \qquad (1)$$

If we know what function in the output space is to be generated for each function in the input space, then $\mathsf{D}$ is completely defined, and, conversely, $\mathsf{D}$ completely defines the behavior of the optical component. The possible input functions $|\phi_I\rangle$ will be in one mathematical function (Hilbert) space – the input space $H_I$ – and the possible output functions $|\phi_O\rangle$ will in general be in another, which we can think of as the output space $H_O$. (See Appendix A for the notation and general properties of these functions and spaces.) We are generally free to choose these spaces (and hence the inputs and outputs we are interested in) to be whatever we want for our device of interest.

The operator $\mathsf{D}$ is similar to the scattering matrix $\mathsf{S}$ encountered with waves or circuits and can be regarded as a general version of $\mathsf{S}$. A key generalization we make for $\mathsf{D}$ is that there is no substantial restriction in our mathematics on what those input or output functions are. More commonly with the $\mathsf{S}$ matrix for circuits or waveguides, the output space is a set of monochromatic modes that are the reflected versions of the input modes. However, we want to be able to consider situations where the input waves are, for example, in one physical volume or surface, and the output waves are in some quite different one. $\mathsf{D}$ also need not be a square matrix (or at least need not start out that way); we could be using a large basis of functions to describe the input wave, such as a plane wave basis, and a limited number of spatial modes to describe the output wave, such as a finite number of modes in a waveguide, for example. D need not be unitary, not only because of possible loss or gain in elements in the device but also because we may physically be scattering into modes that are not included in our mathematical output space – for example, a physically scattered wave might actually miss the output volume.

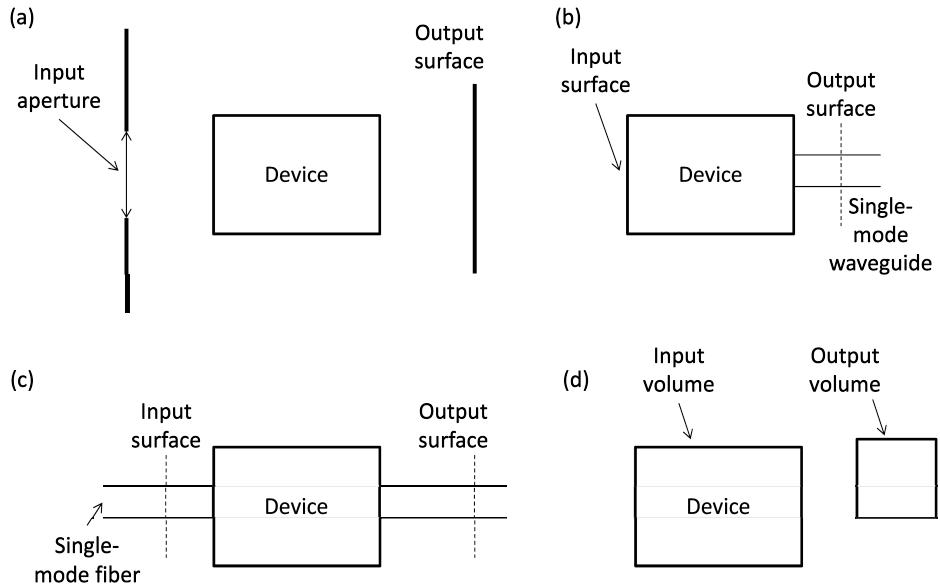

Fig. 1. Illustration of four representative optical device configurations.

Note too that in our mathematics we are explicitly allowing the possibility that our device could contain elements that vary in time (see, e.g., Refs. [8, 10, 11]) and/or space. Of course, though D may represent a device structure whose behavior varies explicitly in time, the mapping it gives between input and output functions is a fixed one. For definiteness of discussion, below we will mostly think of the device as varying only in space, but we understand that variations in time are also covered by the mathematics. In particular, in our way of using the word "mode" here, we include the idea of pulses in time as being modes; we can just as well have pulse modes that are orthogonal in time as we can have spatial modes orthogonal in space, and we can have modes with combinations of spatial and temporal variations.

To preserve the linearity, any such variations in time in the device's properties are predetermined, such as a fixed oscillation in time or some other explicit time dependence of some dielectric constant, and are not dependent on the inputs or output. (Similarly, any variations in real space of the materials are predetermined and are also not dependent on the inputs or outputs of the device.)

## 2.2 Example optical systems

The mathematical approach here is very general, and can describe a broad variety of physical systems. Fig. 1 illustrates example optical systems with input and output spaces. Fig. 1(a) shows a conventional optical system mapping light from one surface to another. The input space is physically waves that are functions of position in the input aperture and the output space is functions of position on the output surface. Fig. 1(b) shows an input coupler for a waveguide. Light focused on the front surface of the device is coupled into a waveguide supporting a single spatial mode. Fig. 1(c) could be an optical fiber dispersion compensator or a filter. The input wave is in a single spatial mode in a single-mode optical fiber, and the mathematical input space is waves as a function of time crossing the input surface in some window of time, and similarly for the output space (possibly with a different window in time). Input functions could be pulses of different center wavelengths, for example, with different delays in each case in the output pulses crossing the output surface, as in some pulse dispersing or dispersion-compensating device, or they could be monochromatic waves, with the device operating as a filter that either transmits or absorbs waves of different incident frequencies. Fig. 1(d) shows a general configuration considering coupling between volumes rather than surfaces. The input physical volume is chosen to coincide with the device volume itself, with the input field being considered as the wave that would have been in the device volume in the absence of the device itself. The output space has a physical volume rather than just being a surface.

Variations of these and many other configurations are also possible. For example, we could extend the configuration of Fig. 1(c) to handle reflective outputs, in which case the output surface might coincide with the input surface, but the mathematical spaces would be different; the input space would be forward propagating waves crossing the that surface, whereas the output space would be backward propagating waves crossing the same surface.

*2.3 Singular value decomposition of the device operator*

The core of our mathematics relies on the singular value decomposition of the operator D. That decomposition is one that allows us to write

$$\mathsf{D} = \sum_m s_{Dm} |\phi_{DOm}\rangle\langle\phi_{DIm}| \quad (2)$$

Equivalently, this means we can write

$$\mathsf{D} = \mathsf{V}\mathsf{D}_{diag}\mathsf{U}^\dagger \quad (3)$$

where U (V) is a unitary operator that in matrix form has the vectors $|\phi_{DIm}\rangle$ ($|\phi_{DOm}\rangle$) as its column vectors and $\mathsf{D}_{diag}$ is a diagonal matrix with diagonal complex number elements $s_{Dm}$.

Here, $s_{Dm}$ are the singular values and the sets of functions $|\phi_{DIm}\rangle$ and $|\phi_{DOm}\rangle$ each form orthogonal sets in their respective spaces $H_I$ and $H_O$. These sets of functions are the solutions of the two eigenvalue problems

$$\mathsf{D}^\dagger\mathsf{D}|\phi_{DIm}\rangle = |s_{Dm}|^2 |\phi_{DIm}\rangle \quad (4)$$

$$\mathsf{D}\mathsf{D}^\dagger|\phi_{DOm}\rangle = |s_{Dm}|^2 |\phi_{DOm}\rangle \quad (5)$$

Note that both $\mathsf{D}^\dagger\mathsf{D}$ and $\mathsf{D}\mathsf{D}^\dagger$ are Hermitian (self-adjoint) operators or matrices even if the matrix D is not, and that these two equations have the same eigenvalues $|s_{Dm}|^2$.

For a very broad range of device operators, which we can reasonably take to include essentially all those of practical interest, this decomposition can be performed. The precise definition of the class of operators for which this is possible is somewhat technical, so we discuss this in detail in Appendix B. The resulting sets of functions – the sets $|\phi_{DIm}\rangle$ and $|\phi_{DOm}\rangle$ for which the singular value $s_{Dm}$ is non-zero – are complete in the following specific sense: Any function in the output space that can be generated by the device from some function in the input space can be written as a linear combination of the set of functions $|\phi_{DOm}\rangle$ corresponding to non-zero singular values, and any function that can be generated by the device in the output space can be generated by some function in the input space that is a linear combination of the $|\phi_{DIm}\rangle$ corresponding to non-zero singular values. In this sense, we will call the sets $|\phi_{DIm}\rangle$ and $|\phi_{DOm}\rangle$ complete orthogonal sets, and we will understand we are including only those associated with non-zero singular values; they certainly can describe all the functions in the input and output space that are of actual interest for the device operation. We will generally take these functions also to be normalized, giving orthonormal basis sets for the spaces of interest.

We now have our general result: Because any linear optical device can be written in terms of a linear operator D between inputs and outputs, and because we can essentially always perform the singular value decomposition of D, then there is a set of orthogonal input functions $|\phi_{DIm}\rangle$ that will give rise, one by one, to a set of corresponding orthogonal output functions $|\phi_{DOm}\rangle$, with non-zero coupling coefficients $s_{Dm}$.

These "mode converter" basis functions $|\phi_{DIm}\rangle$ and $|\phi_{DOm}\rangle$ and values $s_{Dm}$ can always be evaluated given D, and are unique (at least within normalization and phase factors and the usual arbitrariness of orthogonal linear combinations of degenerate eigenfunctions). The resulting device matrix or operator is diagonal when expressed in these basis sets. Hence, any linear optical device can be written as a mode converter, from specific orthogonal input modes to specific orthogonal output modes.

## 3. Derivation of example results

Here we will use the properties of the mode-converter basis $|\phi_{DIm}\rangle$ and $|\phi_{DOm}\rangle$, the corresponding device operator D and the singular values $s_{Dm}$ to solve two particular problems: First, we derive the alignment tolerance of an efficient mode coupler; second, we prove that the loss-less combination of the power from two orthogonal modes is impossible. We are not aware of prior published derivations either of this general expression for alignment tolerance or of a formal proof in terms of modes of the impossibility of such loss-less combination.

*3.1 Mathematical preliminaries*

The following arguments do not require that we are very specific about the kind of device we are considering – the results below are general – but it may be easier for the sake of definiteness to imagine, for example, that we are coupling an optical fiber or focused spot to a small waveguide of some optoelectronic device (e.g., as in Refs. [4, 6, 12]). We will make this coupling with some optical coupler device, which, being a linear optical component, can always be described by a device operator D of the nature we have discussed above. The input functions to this coupler could be waves $|\phi_I\rangle$ that we shine onto the input face of the coupler, with resulting output waves $|\phi_O\rangle$ just inside the waveguide, as in Fig. 1(b).

Now, we can in principle perform the singular value decomposition of the operator D, obtaining the specific orthonormal sets of input and output functions $|\phi_{DIm}\rangle$ and $|\phi_{DOm}\rangle$ with corresponding singular values $s_{Dm}$. As discussed in Appendix A, we can choose the functions to be normalized so that $\langle\phi_{DIm}|\phi_{DIm}\rangle = 1$ and $\langle\phi_{DOm}|\phi_{DOm}\rangle = 1$ correspond to unit power (in the case of steady beams) or energy (in the case of pulses) in each case.

Among the set of values of $s_{Dm}$, there will be at least one with largest possible magnitude, $|s_{max}|$. It is possible that there are several different pairs of functions (and hence values of the index *m*) that all have the same magnitude of singular value, all of those magnitudes being equal to |$s_{max}$|; in such a case, we are free to form any set of orthogonal linear combinations of this subset of input functions (and the same linear combination of this subset of output functions) and use those in the sets |$\phi_{DIm}$⟩ and |$\phi_{DOm}$⟩ at our convenience, as is usual in handling the multiple eigenfunctions of degenerate eigenvalues.

Now, for both our example results here we presume here that we are dealing with a loss-less coupler that is to couple with 100% efficiency from an input mode (or modes) of interest to the output mode of interest (e.g., in the waveguide). Presuming there is no amplification mechanism within the device, then this pair of functions must be one of the possible singular value decomposition pairs with the largest magnitude, $s_{max}$, of singular value; obviously we cannot have a larger singular value than this because that would have to correspond to more than 100% efficiency, which is impossible by definition for our loss-less linear optical component. Because of our choice of power or energy normalization of the basis functions in each space, |$s_{max}$| = |$s_{max}$|$^2$ = 1 for such a 100% efficient coupler. Either (i) the specific mode of interest in the guide and its corresponding input mode already are uniquely the only function pair with this magnitude |$s_{max}$| of singular value, or (ii) we are free to construct linear combinations, as discussed above, such that one of the function pairs with this magnitude |$s_{max}$| of singular value corresponds to our function pair of interest for 100% efficient mode coupling; we can number the singular value functions such that increasing *m* corresponds to progressively smaller magnitudes of singular values, so in either case (i) or (ii) we can choose to call this pair |$\phi_{DI1}$⟩ and |$\phi_{DO1}$⟩.

*3.2 Limit to the alignment tolerance of high-efficiency mode couplers*

Suppose that instead of our ideal input function |$\phi_{DI1}$⟩ we have some other input function |$\phi_{Imis}$⟩, corresponding to some misaligned input. Because we are interested in relative efficiencies of coupling, we take this input function also to have unit power or energy, so ⟨$\phi_{Imis}$|$\phi_{Imis}$⟩ = 1. Now, we can decompose this misaligned input function |$\phi_{Imis}$⟩ into a linear combination of the singular value decomposition set |$\phi_{DIm}$⟩ plus possibly some other function |$\phi_N$⟩ that is orthogonal to all the |$\phi_{DIm}$⟩ (i.e., ⟨$\phi_N$|$\phi_{DIm}$⟩ = 0 for all *m*), giving

$$|\phi_{Imis}\rangle = \sum_m a_m |\phi_{DIm}\rangle + |\phi_N\rangle \qquad (6)$$

where

$$a_m = \langle \phi_{DIm} | \phi_{Imis} \rangle \qquad (7)$$

Here we come to the key point in the argument. Each of the components $a_m$|$\phi_{DIm}$⟩ in the input wave leads to a corresponding wave proportional to |$\phi_{DOm}$⟩. Because any function |$\phi_N$⟩ orthogonal to all the |$\phi_{DIm}$⟩ necessarily leads to no outputs in $H_O$, D|$\phi_N$⟩ = 0, and for all components $a_m$|$\phi_{DIm}$⟩ other than for *m* = 1, the resulting generated waves in the waveguide are orthogonal to our output mode of interest. In other words, none of these other components in the input wave leads to any coupling whatsoever into our desired output mode |$\phi_{DO1}$⟩. We can show this formally by considering the wave generated in the output space (i.e., the waveguide), which is, by definition,

$$|\phi_{Omis}\rangle = \mathsf{D}|\phi_{Imis}\rangle = \sum_m a_m s_{Dm} |\phi_{DOm}\rangle \qquad (8)$$

So, the component of |$\phi_{Omis}$⟩ that is coupled into our mode of interest is

$$\langle \phi_{DO1} | \phi_{Omis} \rangle = a_1 s_{max} \qquad (9)$$

so that the field in our output mode is $a_1 s_{max}$|$\phi_{DO1}$⟩. Since by choice here a field of $s_{max}$|$\phi_{DO1}$⟩ corresponds to unit power transfer efficiency – i.e., ⟨$\phi_{DO1}$|$s^*_{max} s_{max}$|$\phi_{DO1}$⟩ = |$s_{max}$|$^2$ = 1 corresponds to unit power in the output beam – then the power efficiency for coupling into our desired output mode is

$$\eta = |a_1|^2 \equiv \left| \langle \phi_{DI1} | \phi_{Imis} \rangle \right|^2 \qquad (10)$$

which is our general result here for the coupling efficiency of a misaligned beam into a 100% efficient coupler.

For illustration, let us imagine that the desired input beam for coupling into our desired output mode is describable as a simple scalar function of the transverse coordinates *x* and *y*,

$$|\phi_{DI1}\rangle \equiv \phi_{DI1}(x,y) \qquad (11)$$

For the specific case where the "misaligned" field |$\phi_{Imis}$⟩ is merely a displaced version of $\phi_{DI1}(x,y)$, that is, $\phi_{Imis}(x,y) = \phi_{DI1}(x-\Delta x, y-\Delta y)$, then the power coupling efficiency as a function of these displacements Δ*x* and Δ*y* will be

$$\eta(\Delta x, \Delta y) = \left| \iint \phi^*_{DI1}(x,y) \phi_{DI1}(x-\Delta x, y-\Delta y) dx dy \right|^2 \qquad (12)$$

For any mode coupler that is 100% efficient when it is perfectly aligned, $\eta(\Delta x, \Delta y)$ is therefore the alignment tolerance of the power coupling efficiency; at any given displacements Δ*x* and Δ*y*, this is also as big as the power coupling can possibly be. Note that we have not only established a bound on the power coupling efficiency from such a misaligned beam; rather we have shown that, for a mode coupler that is 100% efficient when perfectly aligned, this expression Eq. (12), or, more generally, Eq. (10), *is* the power coupling efficiency when the input beam is misaligned. Note that this expression is now only a function of the input beam shape itself, not of anything else; specifically, it does not depend on the size of the waveguide into which we are coupling.

In our misalignment, we are assuming the mode coupler device itself is perfectly aligned with the output waveguide; this could be the case if these were co-manufactured, as in a taper or inverse taper coupler, for example [7,12]. Alternatively, we could presume that the coupler device was co-manufactured with the input optics, as in, say, a lensed fiber. The mathematics here can equally well be run the other way round in that case, with the misalignment being that of the output waveguide. In that case, we would obtain expressions like Eq. (10) or (12), but with the output beam $|\phi_{DO1}\rangle$ in the expressions instead of the input beam $|\phi_{DI1}\rangle$.

*3.3 Proof of impossibility of loss-less beam combination of multiple modes*

We consider here perfectly loss-less coupling of beams from possibly some number of input modes into one output mode, using an optical device that has no amplification mechanism within it. We presume, then, that we have one pair of perfectly coupled modes, $|\phi_{DI1}\rangle$ and $|\phi_{DO1}\rangle$, as discussed above. Necessarily, these two modes are a pair of the mode-converter basis modes for the device operator D. Suppose, then, that we consider some other input mode $|\phi_{Dextra}\rangle$, orthogonal to $|\phi_{DI1}\rangle$. Since by choice $|\phi_{Dextra}\rangle$ is orthogonal to $|\phi_{DI1}\rangle$, then there is no component of $|\phi_{DI1}\rangle$ in the expansion for $|\phi_{Dextra}\rangle$. $|\phi_{Dextra}\rangle$ is a linear combination of the modes $|\phi_{DIm}\rangle$ for $m \geq 2$ and/or contains functions orthogonal to all the $|\phi_{DIm}\rangle$. Hence, there is no coupling of the power from $|\phi_{Dextra}\rangle$ into the output mode $|\phi_{DO1}\rangle$; instead, because of the one-to-one mapping of the mode-converter basis functions, with input mode $|\phi_{DIm}\rangle$ coupling only into output mode $|\phi_{DOm}\rangle$, any such power is coupled into the other orthogonal modes, $|\phi_{DOm}\rangle$ for $m \geq 2$, or it is not coupled into any of them. Hence, loss-less coupling from two orthogonal modes into one is not possible for any linear optical component (without an amplification mechanism).

This result of the impossibility of loss-less coupling is expected from the Second Law of Thermodynamics; if we could perform such loss-less coupling from two modes into one, then we could make an apparatus that would heat up a warmer black body from the outputs of two cooler black bodies, thus violating the Second Law. Specifically, the apparatus could consist of two black body radiators at low temperatures, each coupled through different single mode output filters, with those two different single modes coupled loss-lessly into a single mode filter at the input to another black body. Thus we could deliver more power into the third black body than is radiated by either of the two cooler black bodies, allowing us to heat up this third black body to a higher temperature than either of the two cool black bodies. Such concepts are discussed in the context also of the Constant Radiance (or Brightness) Theorems [13], though those are usually discussed in the language of imaging optics schemes and are not explicitly stated in term of modes. Our proof here is, however, independent of the Second Law and of the Constant Radiance (or Brightness) Theorems, giving a microscopic argument why such a scheme is not possible for any linear optics.

## 4. Conclusions

We have shown here that any linear optical component can be considered as a mode converter that couples, one by one, from each of a set of orthogonal input modes to each of a set of orthogonal output modes. Thus there is a pair of "mode-converter" basis sets, one for the input and one for the output, and each orthogonal and complete for our problems for every linear optical component. The component can have any specified variation of its properties in space or time. The idea of modes here is understood to refer variations in both space and time of the relevant fields, so it includes both spatial beam forms as well as pulse shapes in general.

We have also illustrated some consequences of this result. Specifically, we have derived a general formula for the alignment tolerance of any 100% efficient mode coupler, and we have proved, without relying on the Second Law of Thermodynamics, that loss-less beam combination of more than one mode into a single mode is impossible. We expect that this approach may enable various other proofs and methods in linear optical components.

## Appendix A – Notation and Hilbert spaces

We are using Dirac notation here to represent the functions, e.g., $|\phi\rangle$, because we want to have a general notation that allows us to consider many different possible kinds of fields; examples of the fields could include monochromatic waves varying in space, vector fields (such as electromagnetic fields), pulsed fields, or conceivably other more complicated fields including other attributes such as quantum mechanical spin. Linear operators are written with *sans serif* upper-case characters, e.g., A. (See, e.g., Ref. [14] for an extended discussion of Dirac notation.)

Generally with the kinds of fields of interest in wave problems, we can define an inner product between functions in each of these spaces. The inner product also allows us to define orthogonality between functions in a space; for two non-zero functions $|\phi_A\rangle$ and $|\phi_B\rangle$ in a given space, $\langle\phi_A|\phi_B\rangle = 0$ means that the functions are orthogonal by definition. With the additional property $\langle\phi_A|\phi_B\rangle = \langle\phi_B|\phi_A\rangle^*$ (where the superscript * denotes the complex conjugate) – a property we typically expect anyway for the overlap integrals of complex wavefunctions of many different types – then essentially we have the conditions for the $H_I$ and $H_O$ to be Hilbert spaces. Since the inner product of a function with itself – i.e., $\langle\phi|\phi\rangle$ – is necessarily real by the above property, we can choose $\langle\phi_I|\phi_I\rangle$ and $\langle\phi_O|\phi_O\rangle$ so that they each represent the power or energy in the field in the input or output space, respectively (though it is not essential that we make that choice).

There is, incidentally, no mathematical requirement that these spaces $H_I$ and $H_O$ are disjoint – they can overlap, with some or all functions of interest being in both spaces, though we will usually think of them as disjoint spaces.

**Appendix B – Conditions for and properties of singular value decomposition**

If an operator is what is known as "compact", the singular value decomposition is always possible (see [15], p. 259, Theorem 4.43). In any reasonable physical problem, we presume the operator D is bounded – that is, it gives a finite output for any finite input. Boundedness is a necessary criterion for an operator D to be compact (see [16], p. 407, Theorem 8.1-3). If we are only considering a finite number of input and output modes for such a bounded operator D, meaning therefore that D is of finite rank, then D is always compact (see [16], p. 407, Theorem 8.1-4).

Compactness covers a very broad range of operators we use with waves even if we are not restricted to a finite number of input and output modes; certainly all the Hilbert-Schmidt operators (see [15], p. 139) are compact (see [15], p. 172). Such Hilbert-Schmidt operators can include ones based on Green's functions for a given wave equation even when those Green's functions themselves are not bounded (Green's functions often are divergent). Certainly if we believe, as we typically do in numerical evaluations, that we can get a sufficiently accurate answer to a given wave problem by using a sufficiently large basis (and hence matrices of sufficiently large size), then we can also perform the singular value decomposition of the underlying operator.

Given that the singular value decomposition can be performed for operators D of interest, we need to understand formally the completeness of the sets $|\phi_{DIm}\rangle$ and $|\phi_{DOm}\rangle$. The mathematical subtlety here is connected with functions associated with singular values $s_{Dm}$ that are zero. In our actual physical problem, we have little or no interest in such functions, since they correspond to inputs that lead to no device output or outputs that cannot be generated by any input. If we choose to work only with functions corresponding to non-zero singular values, then we can avoid these mathematical subtleties, and we have given in the main text above a suitable statement of that restriction and hence of the specific sense of completeness of sets that we use.

We note first of all that, if D is compact, $D^\dagger D$ and $DD^\dagger$ are also compact, and they are also necessarily Hermitian, because they are products of an operator and its Hermitian adjoint. The arguments on completeness of the sets $|\phi_{DIm}\rangle$ and $|\phi_{DOm}\rangle$ are formally slightly different depending on whether D is of finite rank. For finite rank, because of the completeness of the sets of eigenfunctions of finite rank compact Hermitian operators (see [15], p. 250, Theorem 4.38), we can conclude that, for any finite number of input and output modes, the resulting sets of eigenfunctions $|\phi_{DIm}\rangle$ and $|\phi_{DOm}\rangle$ of the finite rank Hermitian operators $D^\dagger D$ and $DD^\dagger$ are then orthogonal and complete for their respective spaces. Even if we are not restricted to finite numbers of input and/or output modes, since $D^\dagger D$ and $DD^\dagger$ are still compact Hermitian operators, by the Hilbert-Schmidt theorem (see [15], p. 257, Theorem 4.40), the resulting eigenfunction sets $|\phi_{DIm}\rangle$ and $|\phi_{DOm}\rangle$ (corresponding to non-zero eigenvalues $|s_{Dm}|^2$) from Eqs. (4) and (5) are complete basis sets in our specific sense. In the main text, therefore, we consider $|\phi_{DIm}\rangle$ and $|\phi_{DOm}\rangle$ to be complete orthogonal sets in our specific sense for their respective spaces $H_I$ and $H_O$.

**Acknowledgments**

This project was supported by funds from Duke University under an award from DARPA InPho program, and by the AFOSR Robust and Complex On-Chip Nanophotonics MURI.